# How to Determine an Optimal Noise Subspace?

Kaijie Xu

*Abstract*—The Multiple Signal Classification (MUSIC) algorithm based on the orthogonality between the signal subspace and noise subspace is one of the most frequently used method in the estimation of Direction Of Arrival (DOA), and its performance of DOA estimation mainly depends on the accuracy of the noise subspace. In the most existing researches, the noise subspace is formed by (defined as) the eigenvectors corresponding to all small eigenvalues of the array output covariance matrix. However, we found that the estimation of DOA through the noise subspace in the traditional formation is not optimal in almost all cases, and using a partial noise subspace can always obtain optimal estimation results. In other words, the subspace spanned by the eigenvectors corresponding to a part of the small eigenvalues is more representative of the noise subspace. We demonstrate this conclusion through a number of experiments. Thus, it seems that which and how many eigenvectors should be selected to form the partial noise subspace would be an interesting issue. In addition, this research poses a much general problem: how to select eigenvectors to determine an optimal noise subspace?

## I. Introduction

Direction Of Arrival (DOA) estimation plays a visible role in array signal processing, and has a wide range of application [1–3]. Multiple Signal Classification (MUSIC) based on the orthogonality between the signal subspace and noise subspace is a most representative algorithm of the high-resolution subspace-based methods [4–6]. Well-known for its high-resolution capability, the MUSIC approach along with their numerous variants can be applied to arbitrary antenna arrays to estimate the DOA of signals, which is regarded as one of the major reasons that it can be widely used in various applications including radar, sonar navigation, mobile communication, and others [7].

High-resolution subspace-based methods estimate the DOA through exploiting the properties of the eigenvectors of the correlation matrix of the received data matrix [8]. The signal and noise subspaces are formed by the eigenvectors corresponding to several (number of signals to be detected) largest eigenvalues and the remaining eigenvectors, respectively. Actually, the MUSIC is essentially a noise-subspace-based algorithm, and its performance of DOA estimation mainly depends on the accuracy of the noise subspace. In the aforementioned researches, the noise subspace is formed by eigenvectors corresponding to all small eigenvalues of the array output covariance matrix [9]. However, after a thorough analysis we found that the estimation of DOA through the noise subspace in the traditional formation is not optimal in almost all cases, and using a partial noise subspace can always obtain better estimation results. In other words, the subspace spanned by a part of the eigenvectors in the traditional noise subspace is more representative of the noise subspace.

The main contribution of this research is to make a thorough analysis of the noise subspace and methods based on it to estimate the DOA of sources, and eventually put forward an ensuing problem: how to select the eigenvectors from the noise eigenvectors to determine an optimal noise subspace?

The study is arranged into five sections. The model of the array signal is formulated in Section II. Section III discusses the noise subspace and puts forward a significant problem. Section IV includes the design of simulation setup and analysis of completed results. Section V covers some conclusions.

## II. Array Signal Model

As shown in Fig. 1, consider $P$ spatial-temporal uncorrelated narrowband far-field signals $\{s_p(t)\}_{p=1}^{P}$ ($t$ indexes the snapshot) [10] with the DOA $\boldsymbol{\theta} = [\theta_1, \theta_2, \cdots, \theta_p, \cdots \theta_P]$ impinging on a uniform linear antenna array (ULA) composed of $M$ omnidirectional antenna elements, where the inter-antenna spacing of each element is $d$. The received signal of the ULA at the $t$th snapshot is given in the following way

$$\boldsymbol{x}(t) = \boldsymbol{A}\boldsymbol{s}(t) + \boldsymbol{n}(t) \quad (1)$$

where $\boldsymbol{A} = [\boldsymbol{a}_1, \boldsymbol{a}_2, \cdots, \boldsymbol{a}_p, \cdots, \boldsymbol{a}_P]$ is the directional matrix (array manifold matrix) and $\boldsymbol{a}_p$ ($p = 1, 2, \cdots, P$) is the $p$th column of $\boldsymbol{A}$, i.e.,

$$\boldsymbol{a}(\theta_p) = \exp\left[1, \cdots, j\frac{2\pi}{\lambda}(m-1)d\sin\theta_p, \cdots, j\frac{2\pi}{\lambda}(M-1)d\sin\theta_p\right]^T \quad (2)$$

$\boldsymbol{s}(t)$ and $\boldsymbol{n}(t)$ respectively denote the signal vector and the noise vector, and $\lambda$ is the wavelength of the signals. Theoretically, the eigenvalue decomposition of the array output covariance matrix is computed in the following way

$$\boldsymbol{R}_x = E\left[\boldsymbol{x}(t)\boldsymbol{x}^H(t)\right] = \boldsymbol{A}\boldsymbol{R}_s\boldsymbol{A}^H + \sigma^2\boldsymbol{I} \quad (3)$$

where $\boldsymbol{R}_s = E[\boldsymbol{s}(t)\boldsymbol{s}^H(t)]$ is the source covariance matrix, and $\sigma^2$ denotes the noise power. The signal and noise subspaces are obtained through the Eigen decomposition of the covariance matrix i.e.,

$$\boldsymbol{R}_x = \boldsymbol{U}_s\boldsymbol{\Lambda}_s\boldsymbol{U}_s^H + \sigma^2\boldsymbol{U}_n\boldsymbol{U}_n^H \quad (4)$$

where $\boldsymbol{\Lambda}_s$ is a diagonal matrix with signal eigenvalues ($P$ largest eigenvalues of $\boldsymbol{R}_x$), and $\boldsymbol{U}_s$ and $\boldsymbol{U}_n$ are the signal and noise subspaces determined by the numerical ordering of the eigenvalues.

K. Xu is with the School of Electronic Engineering, Xidian University, Xi'an 710071, China. (e-mail: kjxu@xidian.edu.cn).

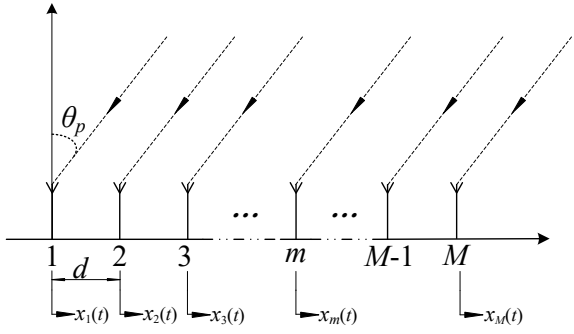

Fig. 1. The modeling scheme of measured data.

Practically, the covariance matrix $R_x$ is unavailable, and it is usually estimated from the collected sample data as [11]

$$\hat{R}_x = \frac{1}{N}\sum_{n=1}^{N} x(t_n)x^H(t_n) \quad (5)$$

where $N$ represents the number of sample data. Then with the signal and noise subspaces various high-resolution methods can be exploited to estimate the DOA.

The MUSIC algorithm estimates the DOA through constructing such a spatial spectrum function [12]

$$F_{\text{MUSIC}}(\theta) = \frac{1}{|a^H(\theta)U_n U_n^H a(\theta)|} \quad (6)$$

On the basis of the orthogonality between the signal subspace and noise subspace, if $\theta$ is a true DOA, the spatial spectrum function will show a large positive value, which is the basic principle of the MUSIC algorithm. Hence, essentially, the MUSIC algorithm is a noise-subspace-based method, and its performance of the DOA estimation mainly depends on the accuracy of the noise subspace.

### III. How to Determine an Optimal Noise Subspace?

In this section, we discuss the noise subspace and its orthogonality with the signal subspace and put forward a significant problem: how to determine an optimal noise subspace?

In the traditional definition, the signal and noise subspaces determined by the distribution of eigenvalues of the array output covariance matrix are defined as $U_s = [u_1, u_2, \cdots, u_i, \cdots, u_P]$ and $U_n = [u_1, u_2, \cdots, u_j, \cdots, u_{M-P}]$, respectively. The orthogonality between the signal subspace and noise subspace also implies that a subspace (partial subspace) spanned by any number of eigenvectors in the noise subspace is orthogonal with the signal subspace.

To facilitate the analysis, we build such a Boolean vector as

$$h = [\cdots, 0, \cdots 1, \cdots] \in R^{1\times(M-P)} \quad (7)$$

with which a partial noise subspace formed by $K$ noise eigenvectors in mathematics can be expressed as

$$\begin{aligned} U_K &= U_n W \\ W &= \text{diag}\{h\} \\ K &= \|h\|_0 \end{aligned} \quad (8)$$

where $\|\cdot\|_0$ denotes $l_0$ norm [13]. By applying the concept of the Weighted MUSIC (WMUSIC) approach [14], here we also consider the matrix $W$ as a weighting matrix for the noise subspace. Through the above analysis, the following orthogonality between the signal subspace and noise subspace is derived

$$U_s \perp U_n \Rightarrow U_s \perp U_K \quad (9)$$

Thus, Eq. (6) can be expressed in the following manner

$$F(\theta) = \frac{1}{|a^H(\theta)U_K U_K^H a(\theta)|} = \frac{1}{|a^H(\theta)U_n W U_n^H a(\theta)|} \quad (10)$$

In other words, when using the MUSIC algorithm to estimate DOA, we do not have to take all the eigenvectors of the noise subspace following the traditional definition. We have also found that the partial subspace can express the noise subspace better than the following traditional definition of the noise subspace, which means that using the partial noise subspace produces better DOA estimation results than that of the traditional manner.

### IV. Experimental Studies

In this section, simulations are presented to evaluate the performance of the developed scheme in comparison with the MUSIC, ESPRIT [15], Root-MUSIC [16] methods. As the most commonly used index, the root mean square error (RMSE) criterion [8] is used in the simulations.

#### A. Simulation Settings

*Simulation 1: RMSE versus SNR.* In the first simulation, we test the RMSE produced by different methods versus varying values of SNR (ranging from -20 dB to 5dB). Consider a 12 elements ULA with relative interelement spacing $d=\lambda/2$, and three uncorrelated narrowband source signals with the DOAs 5°, 10°, 30° impinging on the ULA. The number of snapshots is fixed as 100. In the MUSIC method and the developed scheme, the search step is set as 0.1. A total of 100 independent trials are carried out, and the means of the RMSE results of the different methods versus SNR are displayed in Fig. 2. For the proposed scheme the results presented are the optimal results based on all-possible partial noise subspaces (all the possible combinations of the noise eigenvectors). Furthermore, the corresponding dimensions (include the means and standard deviations) of the partial noise subspace of the proposed scheme are also presented.

*Simulation 2: RMSE versus the number of snapshots.* In the second simulation, we test the RMSE of different methods versus the number of snapshots. The SNR is fixed as -10dB, and the remaining simulation conditions are identical with those used in the first simulation. The RMSE of the different methods versus the number of snapshots (ranging from 20 to 100) is depicted in Fig. 3.

## V. CONCLUSIONS

In this research, we develop an augmented DOA estimation scheme through selecting a collection of the noise eigenvectors to form a partial noise subspace. Compared with the traditional DOA estimation methods, the developed estimator can always achieve better performance than the other estimators in each scenario, and it is also insensitive to low SNRs and small snapshots.

In a nutshell, this research opens a specific way for improving the performance of the DOA estimation and also poses a much general problem: How to select the noise eigenvectors to determine an optimal noise subspace?

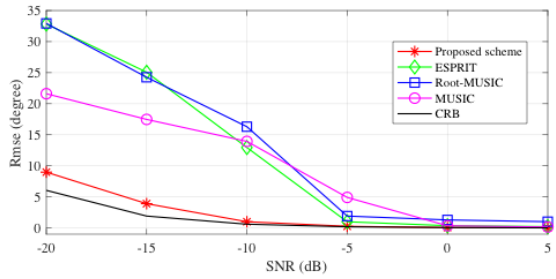
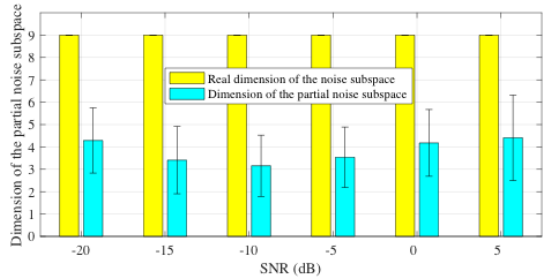

Fig. 2. RMSE versus SNR.

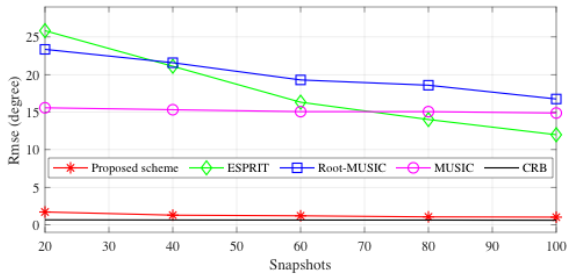
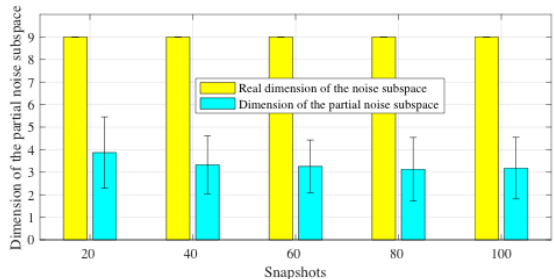

Fig. 3. RMSE versus the number of snapshots.

### B. Discussion and Analysis

It can be seen from Figs. 2 and 3 that the developed scheme achieves better estimation accuracy than that of all the other methods in each scenario, and the performance of the developed scheme is insensitive to low SNRs and small snapshots and is also close to Cramer-Rao Bound (CRB) [17]. In the traditional definition, for three uncorrelated narrowband source signals impinging on 12 elements ULA, the accurate dimension of the noise subspace is nine. However, when we select about three noise eigenvectors to form a noise subspace to estimate DOA, the estimated results are always optimal, which is visualized by the results shown in Figs. 2 and 3.

In summary, using a collection of appropriate noise eigenvectors to replace the traditional definition of noise subspace significantly improves the performance of the high-resolution subspace-based method. At the same time, the sensitivity to low SNR thresholds and small snapshots is also eliminated.

REFERENCES

[1] P. Chen, Z. Cao, Z. Chen and X. Wang, "Off-grid DOA estimation using sparse Bayesian learning in MIMO radar with unknown mutual coupling," *IEEE Transactions on Signal Processing,* vol. 67, no. 1, pp. 208–220, 1 Jan. 2019.

[2] Z. Zheng and S. Mu, "Two-dimensional DOA estimation using two parallel nested arrays," *IEEE Communications Letters,* vol. 24, no. 3, pp. 568–571, Mar. 2020.

[3] Y. Guo, Z. Zhang, Y. Huang and P. Zhang, "DOA estimation method based on cascaded neural network for two closely spaced sources," *IEEE Signal Processing Letters,* vol. 27, pp. 570–574, 2020.

[4] K. Xu, W. Pedrycz, Z. Li and W. Nie, "High-Accuracy Signal Subspace Separation Algorithm Based on Gaussian Kernel Soft Partition," *IEEE Transactions on Industrial Electronics,* vol. 66, no. 1, pp. 491–499, Jan. 2019.

[5] C. Qian, L. Huang, and M. Cao, "PUMA: An improved realization of MODE for DOA estimation," *IEEE Transactions on Aerospace and Electronic Systems,* vol. 53, no. 5, pp. 2128–2139, Oct. 2017.

[6] J. Pan, M. Sun, Y. Wang and X. Zhang, "An enhanced spatial smoothing technique with esprit algorithm for direction of arrival estimation in coherent scenarios," *IEEE Transactions on Signal Processing,* vol. 68, pp. 3635–3643, 2020.

[7] K. J. Xu, W. K. Nie, D. Z. Feng, X. J. Chen, D. Y. Fang, A multi-direction virtual array transformation algorithm for 2D DOA estimation, *Signal Processing,* vol. 125, no. C, pp. 122–133, Aug. 2016.

[8] K. Xu, Y. Quan, B. Bie, M. Xing, W. Nie and H. E, "Fast Direction of arrival estimation for uniform circular arrays with a virtual signal subspace," *IEEE Transactions on Aerospace and Electronic Systems,* doi: 10.1109/TAES.2021.3050667.

[9] X. Li, W. Zhang, T. Shu and J. He, "Two-dimensional direction finding with parallel nested arrays using DOA matrix method," *IEEE Sensors Letters,* vol. 3, no. 7, pp. 1–4, Jul. 2019.

[10] Y. Dong, C. Dong, "Conjugate augmented spatial temporal technique for 2-D DOA estimation with L-shaped array," *IEEE Antennas and Wireless Propagation Letters,* vol. 14, pp. 1622–1625, 2015.

[11] X. Yang, Y. Wang and P. Chargé, "Modified DOA estimation with an unfolded co-prime linear array," *IEEE Communications Letters,* vol. 23, no. 5, pp. 859–862, May 2019.

[12] C. Wang, W. Zheng, P. Gong, "Joint angle and range estimation in the fda-mimo radar: the reduced-dimension root music algorithm, *Wireless Personal Communications,* vol. 118, no. 3, 2515–2533, Jun. 2020.

[13] Z. Shi, T. Shi, M. Zhou and X. Xu, "Collaborative Sparse Hyperspectral Unmixing Using l0 Norm," *IEEE Transactions on Geoscience and Remote Sensing,* vol. 56, no. 9, pp. 5495–5508, Sept. 2018.

[14] P. Stoica and A. Nehorai, "MUSIC, maximum likelihood, and Cramer-Rao bound," in IEEE Transactions on Acoustics," *Speech, and Signal Processing,* vol. 37, no. 5, pp. 720–741, May 1989.

[15] R. ROY, "ESPRIT-estimation of signal parameters via rotational invariance techniques," *IEEE Transactions on Acoustics, Speech, and Signal Processing,* vol. 37, no. 7, pp. 984–995, Jul. 1989.

[16] M. Wagner, Y. Park and P. Gerstoft, "Gridless DOA Estimation and Root-MUSIC for Non-Uniform Linear Arrays," *IEEE Transactions on Signal Processing,* vol. 69, pp. 2144–2157, 2021.

[17] M. Trinh-Hoang, M. Viberg and M. Pesavento, "Cramer-Rao Bound for DOA Estimators Under the Partial Relaxation Framework: Derivation and Comparison," *IEEE Transactions on Signal Processing,* vol. 68, pp. 3194–3208, May 2020.